\documentclass[reprint,aps,prl]{revtex4-1}

\usepackage{amssymb}
\usepackage{amsfonts}
\usepackage{amssymb}
\usepackage{amsmath}
\usepackage[dvips]{graphicx}
\usepackage{color}
\usepackage{ulem}


\begin{document}

\title{High-Efficiency Quantum Memory of Full-Bandwidth Squeezed Light}
\author{Jinxian Guo$^{1}$}
\author{Meihong Wang$^{2,3}$, Zeliang Wu$^{4}$, Chenyu Qiao$^{2,3}$, Fengyi Xu$^{2,3}$, Xiaoran Zhang$^{2,3}$,  Xiaolong Su$^{2,3}$}
\email{suxl@sxu.edu.cn}
\author{Liqing Chen$^{3,6}$}
\email{lqchen@phy.ecnu.edu.cn}
\author{Weiping Zhang$^{1,3,5,6}$}
\email{wpz@sjtu.edu.cn}

\affiliation{$^{1}$ School of Physics and Astronomy, and Tsung-Dao Lee Institute,  Shanghai Jiao Tong University, Shanghai, 200240, China\\
$^{2}$ State Key Laboratory of Quantum Optics Technologies and Devices, Institute of Opto-Electronics, Shanxi University, Taiyuan 030006, China\\
$^{3}$ Collaborative Innovation Center of Extreme Optics, Shanxi University, Taiyuan, 030006, China\\
$^{4}$ School of Physics and Electronic Science, East China Normal University, Shanghai 200062, China\\
$^{5}$ Shanghai Research Center for Quantum Sciences, Shanghai, 201315, China\\
$^{6}$ Shanghai Branch, Hefei National Laboratory, Shanghai, 201315, China}

\begin{abstract}
In continuous-variable quantum information processing, it is crucial to develop high-efficiency and broadband quantum memory of squeezed light, which enables the memory of full-bandwidth information.
Here, we present a quantum memory of squeezed light with up to 24 MHz bandwidth, which is at least 12 times that of previous narrowband resonant memory systems, via a far-off resonant Raman process. 
We achieve output squeezing of as high as 1.0 dB with fidelity above 92\% and a memory efficiency of 80\%, corresponding to an end-to-end efficiency of 64.2\%, when input squeezing is 1.6 dB.
The lowest excess noise of 0.025 shot-noise-unit in the memory system is estimated by the noisy channel model which is benefited from optimizing quantum memory performance with a backward retrieval strategy.
Our results represent a breakthrough in high-performance memory for squeezed states within tens of MHz-level bandwidth, which has potential applications in high-speed quantum information processing.
\end{abstract}

\pacs{Valid PACS appear here}
\maketitle

\affiliation{$^{1}$School of Physics and Astronomy, and Tsung-Dao Lee Institute,
Shanghai Jiao Tong University, Shanghai 200240, P. R. China ~\\
 $^{2}$State Key Laboratory of Precision Spectroscopy, School of
Physics and Electronics, East China Normal University, Shanghai 200062,
P. R. China~\\
 $^{3}$Shanghai Research Center for Quantum Sciences, Shanghai 201315,
P. R. China~\\
 $^{4}$Collaborative Innovation Center of Extreme Optics, Shanxi
University, Taiyuan, Shanxi 030006, P. R. China~\\
 $^{5}$Shanghai Branch, Hefei National Laboratory, Shanghai 201315,
P. R. China}

\textit{Introduction.}-- As one of the main approaches to quantum information processing, continuous-variable (CV) quantum information has the unique advantages of encoding in an infinite-dimensional Hilbert space and enabling unconditional information processing \cite{braunstein_quantum_2005, RevModPhys.84.621, WANG20071}.
Significant advances have been achieved in CV quantum information including unconditional quantum teleportation \citep{huo_deterministic_2018}, quantum memory \citep{honda_storage_2008, appel_quantum_2008, jensen_quantum_2011}, quantum key distribution \citep{jain_practical_2022, hajomer_long-distance_2024}, and quantum computing \citep{konno_logical_2024, enomoto_continuous-variable_2023}.
Squeezed light, whose noise of one quadrature is below the shot noise limit, is an essential quantum resource for CV quantum information and quantum-enhanced measurements \citep{braunstein_quantum_2005, Andersen_2016}. 
As technology advances, it is essential to develop high-performance broadband quantum memory for squeezed light \citep{kuruma_controlling_2024, liu_quantum_2023}, which is a key element in scalable and high-speed CV quantum information \citep{mol_quantum_2023, hillmann_universal_2020}.

Until now, the demonstrated memories of squeezed light only present a MHz-level bandwidth in narrow-bandwidth memory systems, including electromagnetically induced transparency (EIT) \citep{honda_storage_2008, appel_quantum_2008, yan_establishing_2017, arikawa_quantum_2010} or quantum non-demolition (QND) measurement of spins \citep{jensen_quantum_2011}. 
The inherent narrow bandwidth of these systems hinders the memory of the broadband squeezed light, leading to the loss of high-frequency information of quantum states. 
In addition to bandwidth, the efficiency and low noise of a quantum memory system are the other two important parameters that affect the memory of full-bandwidth squeezed light. 
Consequently, the development of high-efficiency, low-noise, and broadband quantum memory is crucial for the practical memory of broadband squeezed state, which is a fundamental requirement for high-fidelity continuous-variable quantum computing and communication.

Far-off resonance Raman quantum memory promises to store broadband quantum states which avoids the imperfections in quantum memory of broadband squeezed light.
Up to now, the memory of weak coherent pulse and single-photon states with a bandwidth of GHz \citep{reim_towards_2010, fisher_frequency_2016} has been demonstrated. 
Recently, the efficiency of far-off resonance Raman quantum memory has been improved by employing optimization techniques \citep{guo_high-performance_2019}. 
Additional noise can also be suppressed by noise suppression methods \citep{England2015Storage, Saunders2016Cavity, Zhang2014Suppression, Thomas2019Raman, Michal2014Hamiltonian, yu2023Interferometry}.
However, quantum memory for squeezed states imposes far stricter end-to-end efficiency and noise control requirements than those demonstrated for coherent pulses or single photons, presenting significant challenges for memory system performance. 
To date, high-efficiency quantum memory for broadband squeezed light remains unrealized.

\begin{figure*}[tbp]
\centering 
\includegraphics{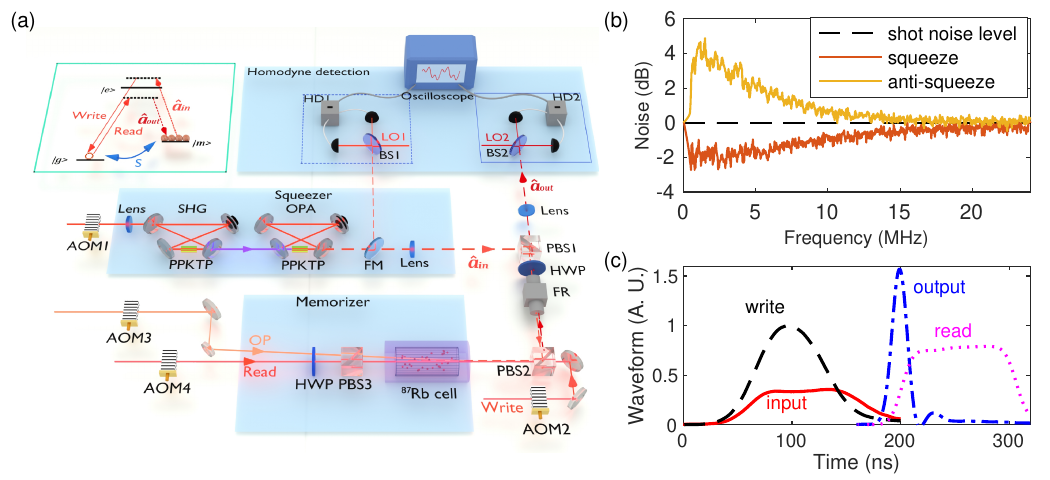}
\caption{(a) Experimental setup of quantum memory. AOM, acousto-optical modulator; SHG, second harmonic generator; OPA, optical parametric amplifier; PPKTP, periodically poled KTiOPO$_{4}$; FM, filp mirror; OP, optical pumping; FR, Faraday rotator; HWP, half-wave plate; PBS, polarized beam splitter; BS, beam splitter;  LO, local oscillator; HD, homodyne detector. The inset at left corner shows the atomic level configuration. The atomic vapor is heated to 89${^{\circ }}C$. The power of write and read lights are 240 mW and 13 mW, respectively. (b) The relative noise power spectrum of a CW squeezed vacuum state. The squeezing sidebands expands to 15 MHz. (c) Temporal waveforms of the input squeezed vacuum state ($\hat{a}_{in}$), the retrieved squeezed vacuum pulse ($\hat{a}_{out}$), the write pulse, and the read pulse in experiment.}
\label{setup}
\end{figure*}
In this work, we demonstrate the write-in and retrieval of a full-bandwidth squeezed vacuum state using a Raman memory in an atomic vapor. 
We develop a dual-pulse time-domain homodyne tomography technique and a backward retrieval strategy to solve the technical challenges of quantum state characterization and quantum memory performance optimization, respectively.
The memory efficiency of 80\% and the excess noise of 0.025 shot-noise-unit (SNU) are achieved by a backward retrieval strategy, which is estimated by a noisy channel model. 
For an input squeezed light pulse with a noise reduction of 1.6 dB, the retrieved state maintains a noise reduction of 1.0 dB, achieving the highest level of squeeze for a retrieved pulse. 
Furthermore, we also demonstrate the quantum memory of a full-bandwidth squeezed state with a maximum bandwidth of 24 MHz.
Benefited from the high-efficiency and low-noise memory system, the fidelity of the retrieved signal exceeds 92\% even when operating at a bandwidth more than an order of magnitude larger than previously reported systems. 
These results represent a significant advancement toward the realization of high-performance continuous-variable (CV) quantum technologies, including quantum computing, squeezing-enhanced quantum metrology, and quantum communication.

\textit{Experimental setup.}-- As shown in Fig. \ref{setup} (a), the experimental setup of broadband squeezed light quantum memory consists of three parts, which are squeezer, memorizer and homodyne detection system, respectively. 
In the part of the squeezer, an optical parametric amplifier (OPA) is used to generate the squeezed vacuum state.
A 795 nm laser with the blue-detuned 1.6 GHz frequency from the $^{87}Rb$ atomic transition $|5S_{1/2},F=2\rangle \rightarrow |5P_{1/2},F^{\prime }=2\rangle$, passes through an acousto-optic modulator (AOM1) to produce the pulsed light with full width at half maxima (FWHM) of 227.2 ns. 
The pulsed light is seeded into a second-harmonic generator (SHG) to generate a pulsed beam at 397.5 nm, which serves as the pump beam of the OPA.
The detailed parameters of SHG and OPA can be found in the supplemental materials \cite{supp}.
Under the de-amplification condition of OPA \citep{wang_experimental_2022} with a pump power of 25 mW, an amplitude-squeezed state is generated with a squeezing bandwidth of 15 MHz. 
As shown in Fig. \ref{setup} (b), the noise of amplitude quadrature ($\hat{x}=\left( \hat{a}+\hat{a}^{\dagger }\right) /\sqrt{2}$) and phase quadrature ($\hat{p}=\left( \hat{a}-\hat{a}^{\dagger }\right) /\sqrt{2}i$) of the squeezed state are squeezed and anti-squeezed respectively, where $\hat{a}^{\dagger}$ and $\hat{a}$ are the creation and annihilation operators.

In the part of the memorizer, a $^{87}Rb$ atomic ensemble is initially prepared in the state $|5S_{1/2}, F=2\rangle$ by a 780 nm pumping laser with FWHM of 45 $\mu s$. 
The squeezed vacuum state $\hat{a}_{in}$ is co-injected with a write pulse into the atomic ensemble to perform the Raman write process, when the flip mirror (FM) is flipped down. 
The write pulse shaped by the AOM2 is two-photon resonant with $\hat{a} _{in}$ in the atomic ensemble, as illustrated in the inset of Fig. \ref{setup} (a). 
After the write process, a read pulse with FWHM of 105 ns, which is shaped by AOM4 with the red-detuned frequency of 2 GHz from the atomic transition $|5S_{1/2},F=1\rangle \rightarrow |5P_{1/2},F^{\prime }=1\rangle $, is injected into the atomic ensemble to retrieve the squeezed vacuum state $\hat{a}_{out}$. Figure \ref{setup} (c) shows the temporal waveforms of the input state, write, read, and output state of the memory process. It can be seen that the bandwidth of the output state reaches 43.1 MHz, which is high enough for the memory of full-bandwidth squeezed state.

In the experiment, we solve two technical challenges, which are characterization of the quantum properties of the broadband squeezed states and achieving a high-efficiency, low-noise Raman quantum memory, respectively, to achieve the quantum memory of the broadband squeezed state.
To solve the first challenge, we develop a dual-pulse time-domain homodyne tomography technique to obtain the temporal waveform and phase of a weak signal, which is difficult in conventional time-domain homodyne tomography \citep{appel_quantum_2008} due to low photon number of the measured quantum state ($<$0.1 photons/pulse).
In our method, we apply the squeezed vacuum state as the first pulse to obtain the quadrature value, and a bright coherent state, which has the same waveform as the first one but delayed by 500 ns, as the second pulse to track the phase.
To reconstruct the input state $\hat{a}_{in}$, we interfere the first pulse with the local oscillator (LO1) on HD1 to acquire the homodyne photoncurrent, a temporal waveform of the input state is obtained through a point-wise calculation of the variance of the homodyne photon current. 
The quadrature value is obtained by integrating the production of the photoncurrent and the temporal waveform \citep{appel_quantum_2008}.
The phase of the squeezed vacuum state is determined by acquiring the interference fringe of the second pulse and the LO1 on HD1.
Since phase drifts typically occur at a low frequency ($<$1 MHz), the effect of delay on phase estimation can be ignored.
To reconstruct the output state $\hat{a}_{out}$, the pulsed squeezed vacuum state is stored and retrieved in the atomic ensemble, followed by the write-in and retrieval of the delayed bright coherent light.
The homodyne photoncurrent and interference phase are obtained by seeding the retrieved squeezed vacuum state and coherent light into HD2, respectively.
Since the phase of the retrieved state is tracked according to the phase of retrieved coherent light, the systematic Stark shifts introduced by quantum memory can be canceled.

To solve the second technical challenge of achieving a high-efficiency, low-noise Raman quantum memory, we apply a backward-retrieval Raman memory strategy.
In the write process, we control the waveform of the write pulse to achieve a precise temporal mode-matching between the input signal and the write pulse which improves the write-in efficiency \citep{guo_high-performance_2019}. 
In the experiment, we use a strong coherent probe with the same temporal envelope as the squeezed vacuum and iteratively update the temporal waveform of the write pulse using a Differential Evolution (DE) algorithm to maximize write-in efficiency \cite{ming_optimizing_2023,supp}. 
In the read process, we use a backward-propagating read light (as shown in Fig. \ref{setup} (a)) to extract the signal out of the atomic ensemble. Compared to the conventional forward retrieval strategy \citep{guo_high-performance_2019}, the backward retrieval strategy achieves higher efficiency with lower read-light power, resulting in lower system noise \citep{supp}. Thus, the loss and FWM noise amplification is reduced by the backward retrieval strategy.

\begin{figure}[tbp]
\centering 
\includegraphics{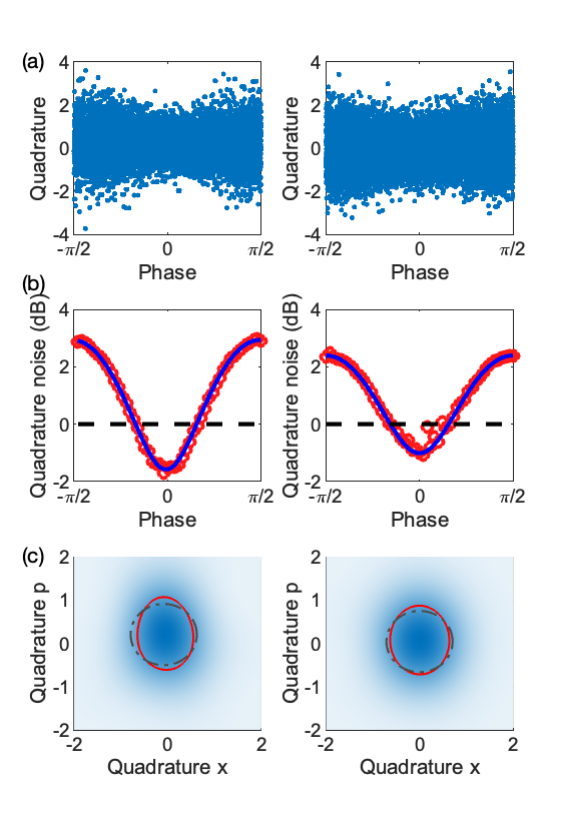}
\caption{(a) Experimental data of phase-dependent quadrature values. Left and right columns represent the input and retrieved states, respectively. (b) Results of quadrature variances. The red circles represent the variances obtained from experimental quadrature values and the solid lines are the corresponding fittings. (c) Contour plots of Wigner functions. The black circles represent the vacuum state and the red circles are the squeezed vacuum states.}
\label{best}
\end{figure}

\textit{Memory Performance.}-- We evaluate the performance of quantum memory by analyzing the quadratures and Wigner functions of the input and output states, as shown in Fig. \ref{best}.
The measured quadrature values of the input and output states are shown in Fig. \ref{best} (a) when the FWHM of the input state is 227.5 ns. 
The noise fluctuations of the amplitude quadrature (phase $0$) of the states $\hat{a}_{in}$ and $\hat{a}_{out}$ are narrower than those of the shot noise level and the amplitude quadrature (phase $\pi/2$), which means that the input and output states of the quantum memory are both squeezed states.
Based on the quadrature data, we obtain the quadrature variances of the input and output states, respectively, as shown in Fig. \ref{best} (b). 
The input state's minimum quadrature noise is 1.6 dB lower than the shot noise level (SNL, black dashed line), whereas the output state remains 1.0 dB below the SNL. 
In the current system, the end-to-end efficiency is 64.2\%, including the memory efficiency of 80\% and the optical transmission efficiency of 80.3\%.
Using the quadrature data, we reconstruct the Wigner functions of the input and output states via maximum likelihood estimation for quantum state tomography \citep{lvovsky_iterative_2004}, as shown in Fig. \ref{best} (c).
Based on the Wigner functions, we obtain the unconditional fidelity $F=Tr[(\rho_{in}^{1/2}\rho_{out}\rho_{in}^{1/2})^{1/2}]^2$ of the memory system by calculating the overlap between the input density matrix $\rho_{in}$ and the output states $\rho_{out}$, which reaches 97.9\%.
These results confirm that our memory effectively preserves the quantum properties of the input state. 

Since the performance of quantum memory is also influenced by the noise level in the memory process, it is important to characterize the excess noise of the memory system. 
Here, we characterize excess noise by introducing a noisy channel model. In this model, the squeezed vacuum state passing through the noisy channel is described by \citep{supp,zhang_one-sided_2025}
\begin{eqnarray}
\hat{a}_{out} = \sqrt{\eta}\hat{a}_{in}+\sqrt{1-\eta}\hat{v}+\hat{b}_{th},\label{BS}
\end{eqnarray}
where $\eta$ is the transmission efficiency of the memory, $\hat{b}_{th}$ and $\hat{v}$ are the operators of thermal and vacuum states, respectively. The excess noise is introduced independently since it acts as a non-degenerate amplification which is mainly due to the FWM process in Raman memory \citep{supp}.
Based on the measured variances of the input and output quadratures, we estimate the excess noise in the memory process by the variance of the thermal state $\hat{b}_{th}$ according to Eq. (\ref{BS}), which is as low as $\delta=0.025$ SNU under the condition that the FWHM of the input state is 227.5 ns. 

\begin{figure}[h]
\centering \includegraphics{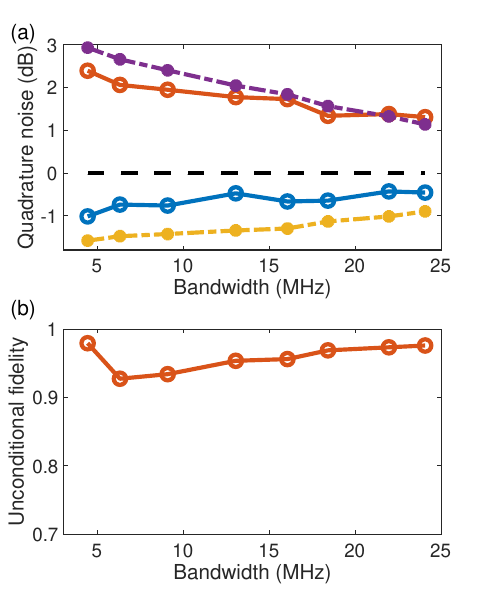}
\caption{(a) The dependence of squeezing level of input and retrieved states on the bandwidth of input state. The yellow (blue) and purple (red) data points represent the squeezing and anti-squeezing of input (retrieved) states. (b) The dependence of unconditional fidelity of the memory process on the bandwidth of input state.}
\label{squeezeband}
\end{figure}

An important advantage of Raman memory is that it enables one to store the signal with a GHz-level bandwidth.
To showcase its broadband capability, we store squeezed states with different bandwidths which are prepared by reducing the pump pulse duration for SHG.
In this case, the squeezed vacuum states $\hat{a}_{in}$ with FWHM varying from 227.2 to 41.6 ns (corresponding to a bandwidth of 4.4 to 24 MHz) are prepared, whose squeezing level varies from 1.6 dB to 0.9 dB, as shown in Fig. \ref{squeezeband} (a). 
After quantum memory, the retrieved states consistently exhibit noise levels below the SNL, which covers the full bandwidth of the squeezed vacuum state. 
At 24 MHz bandwidth, the output state retains 0.55 dB of squeezing despite the input signal's 0.9 dB squeezing, corresponding to a memory fidelity of 97.6\%.
The 24 MHz bandwidth is 12 times larger than the previous EIT-based system.
To better demonstrate the broadband capability of the memory system, we further test the unconditional fidelity change with input signal bandwidth, as shown in Fig. \ref{squeezeband} (b).  
The unconditional fidelity all exceeds 92\% in the bandwidth range from 4.4 to 24 MHz. 

\renewcommand{\arraystretch}{2}
\begin{table}[htbp]
 \caption{\label{excess}Parameters of bandwidth $B$ (MHz), the end-to-end efficiency $\eta$ and the excess noise $\delta$ (SNU) in the memory process.}
 \setlength{\tabcolsep}{1.5mm}
  \begin{tabular}{@{}lllllllll@{}}
    \hline
    \hline
  \multicolumn{1}{c}{$B$}&4.4&6.28&9.06&13.0&16.0&18.4&21.9&24.0 \\
    \hline
    \multicolumn{1}{c}{$\eta$} &0.642&0.635    &0.666&0.630&0.742&0.629&0.663&0.757\\
    \hline
    \multicolumn{1}{c}{$\delta$} &0.025&0.026& 0.023&0.047&0.038&0.013&0.025&0.021 \\
    \hline
     \hline
  \end{tabular}
\end{table}

Furthermore, we evaluate end-to-end efficiency $\eta$ and excess noise $\delta$ for input states with different bandwidths, as shown in Table \ref{excess}. The end-to-end efficiencies do not decrease significantly with the increase of bandwidth and excess noises are typically around 0.027 SNU, which ensures a high squeezing level of the retrieved state.
These results demonstrate that current memory is capable of storing a full-bandwidth squeezed state.

\textit{Discussion and Conclusion.}-- In summary, we experimentally demonstrate a high-performance memory of full-bandwidth squeezed vacuum states with Raman quantum memory. By applying a backward retrieval strategy, we realize a Raman quantum memory with memory efficiency of 80\% and excess noise of 0.025 SNU, which provides a high-efficiency and low-noise quantum memory of squeezed state. 
Utilizing a dual-pulse time-domain homodyne tomography, we obtain the retrieval state with a maximum squeezing of 1.0 dB when a squeezed state with only 1.6 dB squeezing is applied as input state. 
To further highlight the broadband capability of current Raman memory, we expand the bandwidth of input states to 24 MHz and achieve 0.55 dB output squeezing, when the input squeezing is 0.9 dB. 
All unconditional fidelities for squeezed states with different bandwidths exceed 92\%, which confirms the successful memory of full-bandwidth squeezed state.

The high efficiency of Raman memory protects the state from loss, and its low noise property protects the state from the contamination of additional noise. 
The performance of the memory system can be further optimized by reducing end-to-end losses and excess noise.
Additionally, the broadband nature of Raman memory enables a higher bandwidth of the retrieved state than that of the input signal, since the bandwidth of the retrieved state is determined by the read pulse \citep{guo_high-performance_2019}. 
In this experiment, a 4.4 MHz squeezed vacuum state input yields a 43.1 MHz squeezed vacuum state output, representing a significant bandwidth enhancement. 
This provides a novel approach for high-performance quantum state conversion interfaces, which has potential applications in continuous-variable (CV) quantum information processing.

We acknowledge financial support from the Innovation Program for Quantum Science and Technology 2021ZD0303200, the National Science Foundation of China (Grant NO. 12434015, U23A2075, 12274132, 12234014, 12374331, 62275145, 11654005), Shanghai Municipal Science and Technology Major Project (Grant NO. 2019SHZDZX01), Innovation Program of Shanghai Municipal Education Commission (No.202101070008E00099), and the National Key Research and Development Program of China under Grant number 2016YFA0302001. W.Z. also acknowledges additional support from the Shanghai talent program.

\appendix
\newpage
\begin{widetext}
\section{S1. Experimental Setup}
The optical layout of our experiment is shown in Fig. \ref{setup}. To ensure phase stability throughout the system, we employed only one master laser at 795 nm. The center frequency of the master laser is set to the signal frequency in the main text. The master laser is first cleaned by a mode-cleaning cavity, then split into three branches: one each for the squeezer, memorizer, and homodyne detection.
\begin{figure}[!]
\centering
\includegraphics{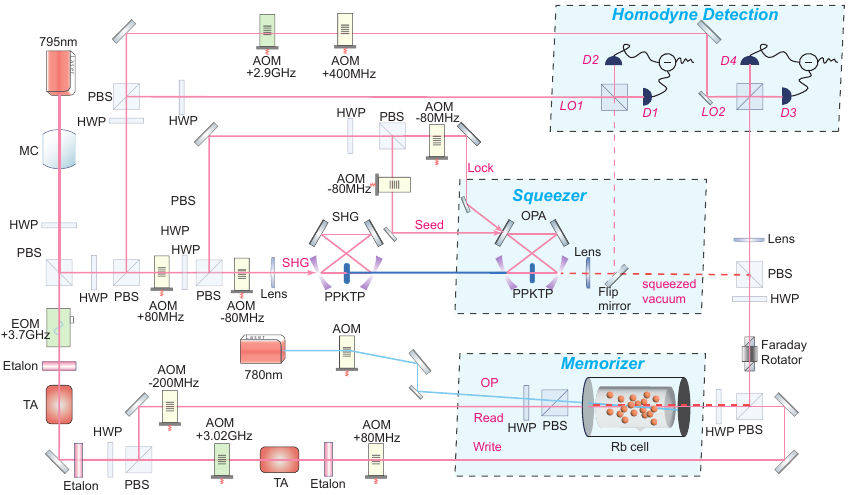}
\caption{\label{setup}The detailed experimental setup. TA: the tapered amplifier; PBS: polarized beam splitter; HWP: half wave plate; AOM: Acoustic optical modulator; EOM: electrical optical modulator; MC: mode cleaner.}
\end{figure}

In squeezer, the master laser is further split into three paths, each controlled by AOM pairs for the pulse control without any frequency shifts. One of the beams is sent into the SHG cavity to generate the 397.5 nm pump light via a PPKTP (1 mm $\times$ 2 mm $\times$ 10 mm) crystal. The SHG cavity has a cavity length of 480 mm, which contains two concave mirrors (R=50 mm), a high reflectivity mirror, a plane mirror with transmissivity of 8\%. The generated 397.5 nm pump light is then sent into the OPA cavity with a cavity length of 480 mm, which contains two
concave mirrors (R=50 mm), a high reflectivity mirror, a
plane mirror with transmissivity of 12.5\% and a PPKTP crystal. The OPA generates a squeezed vacuum state by setting the cavity on parametric amplification status, which is guaranteed by locking the relative phase between the pump beam and the signal beam to zero. The locking laser and the seed light are injected into the cavity from the plane mirror. 

In the memorizer, the master laser is modulated by an EOM driven by a microwave generator with frequency 3.7 GHz. The positive sideband is filtered out by an etalon with FSR of 10 GHz and transmission FWHM of 1 GHz. The filtered beam is amplified using a tapered amplifier (TA), and a second etalon is applied after the amplification to suppress the fluorescence noise. Then the write light is frequency-shifted by an AOM of 3.02GHz, followed by a second TA and another etalon. After the etalon, the write light is chopped into pulses. The read beam is derived from the 3.7 GHz shifted master laser, chopped by a 200 MHz AOM. The 780 nm optical pumping beam for the atomic ensemble is pulsed using an 80 MHz AOM.

In the homodyne detection setup, two local oscillators (LOs) are required. LO1, resonant with the input squeezed vacuum, is directly came from the master laser. The relative phase between the LO1 and the squeezed vacuum state is scanned by a piezoelectric transducer (PZT) driven mirror. LO2, resonant with the retrieved signal, is generated by first shifting the master laser through a 2.9 GHz AOM and then a 400 MHz AOM, achieving the desired offset. The relative phase between the LO2 and the retrieved state is scanned by shifting the 400 MHz AOM by 100 Hz. The bandwidth of the homodyne detector is 45 MHz and the detection efficiency includes the interference efficiency between signal light and local light (98.5\% and 95\% for the input and output states, respectively) and the quantum efficiency of the photodiodes (92\%, S3883). 

\section{S2. Dual-pulse Time-domain Homodyne Tomography}

To characterize the quantum properties of the squeezed vacuum pulses before and after memory memory, we developed a dual-pulse time-resolved homodyne detection scheme. The detailed pulse sequence is depicted in Fig. \ref{time}.
\begin{figure}[!]
\centering
\includegraphics[scale=0.85]{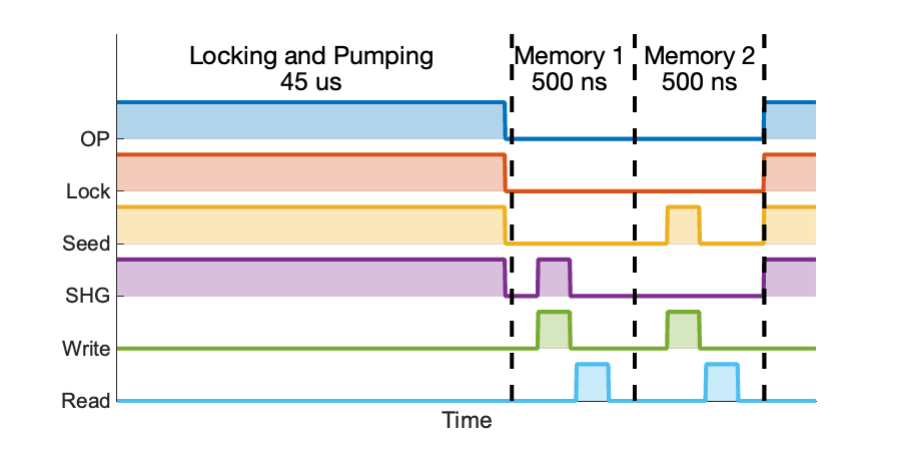}
\caption{\label{time}The time sequence of the pulses. OP: the 780 nm optical pumping for atomic ensemble; Lock: the locking laser for OPA cavity; Seed: the phase locking laser for squeezed vacuum state; SHG: the optical pumping and locking laser for SHG cavity to generate 397.5 nm; Write: the write pulse to write the state into atomic ensemble; Read: the read pulse to read the state out from atomic ensemble.}
\end{figure}

In the beginning of the pulse sequence cycle, the two locking pulses for the cavity, the seed pulse and the 780 nm pumping pulse are turned on, which last for 45 $\mu s$. After the locking pulses, the first memory process begins. The pumping light and the write pulse are switched on, generating a squeezed vacuum field which is stored in the atomic ensemble. These pulses are then turned off, and the read pulse is applied to retrieve the quantum state, which is sent to the HD for quadrature measurement.

Following the first memory sequence, a second memory process is performed. The second process starts with a strong coherent pulse from the seed light with the same temporal envelope as the squeezed vacuum state. This coherent pulse is stored using a second write pulse with the same waveform in the first process, then retrieved using a read pulse and directed to HD. The two memory events are separated by only 500 ns, and with a typical optical phase drift below 100 kHz, the phase drift between them is negligible.

In the first memory process, we obtain homodyne-measured photocurrents for both the input and the output quantum states. In the second memory process, we extract interference fringes $f(t)$ by integrating the homodyne photocurrent of the coherent states. Applying a Hilbert transform to the interference fringes, we recover the relative phase as
$\phi(t) = tan^{-1}\left(\text{Im}\{H[f(t)]\}/f(t)\right)$, where $\text{Im}\{H[f(t)]\}$ represents the imaginary part of the Hilbert transform of $f(t)$.

\section{S3. Backward Retrieval Strategy}

To achieve high-efficiency and low-noise quantum memory, we employ a backward retrieval strategy, which is modeled using noisy Raman memory equations in co-moving frame \cite{Thomas2019Raman},
\begin{eqnarray}
\frac{\partial}{\partial z}\hat{a}_{s}&=&g_{s}\Omega(t)\hat{s}\\
\frac{\partial}{\partial z}\hat{a}_{a}^{\dagger}&=&g_{a}\Omega(t)\hat{s}e^{i\Delta k z}\\
\frac{\partial}{\partial t}\hat{s}&=&g_{s}\Omega^*(t)\hat{a}_{s}-g_{a}\Omega^*(t)\hat{a}_{a}^{\dagger}e^{-i\Delta k z}
\end{eqnarray}
where $\hat{a}_{s}$, $\hat{a}_{a}$, $\hat{s}$ are signal, amplified anti-Stokes and spin wave, respectively. $g_{s}$ and $g_{a}$ are the coupling coefficients for signal-spin and anti-Stokes-spin coupling, respectively. $\Omega(t)$ is the Rabi frequency of write/read light. 

In the writing process, the optical signal $\hat{a}_{s}$ is mapped onto an atomic spin-wave $\hat{s}$ which is controlled by the write pulse. Efficient writing requires precise temporal mode-matching between the input signal and the write pulse \cite{guo_high-performance_2019}. To optimize this, we use a strong coherent probe with the same temporal envelope as the squeezed vacuum and iteratively update the temporal waveform of the write pulse using a Differential Evolution (DE) algorithm to maximize storage efficiency \cite{ming_optimizing_2023}. 

During optimization, a population of Gaussian pulses is initially generated, each characterized by a randomly assigned temporal delay ($\tau_0$) and full width at half maximum (FWHM, $\Delta\tau$). These parameters define the temporal waveform of each write pulse. The fitness of each pulse is evaluated by modulating the AOM of write light with the temporal waveform, with the storage efficiency serving as the fitness function. The population is then evolved over successive generations via standard genetic operations, including mutation (random perturbation of $\tau_0$ and $\Delta\tau$), crossover (recombination of parameter pairs), and selection (preferential retention of high-efficiency pulses). This optimization process continues iteratively until the storage efficiency converged to a global or sufficiently high local maximum. The resulting set of parameters correspond to the optimal temporal shape of the write pulse, tailored to maximize the efficiency of quantum state storage in the atomic ensemble.

After optimization of the write process, we evaluate the memory efficiencies for both forward and backward retrieval strategies (see Fig. \ref{read} (a)). The results show that forward retrieval requires higher readout power for higher efficiency. This also induced an increase of excess noise. Figure \ref{read} (b) shows that the excess noise increases with the read power. In contrast, backward retrieval achieves high efficiency at a low power which gives a low noise, making it preferable for the quantum memory of squeezed vacuum states.
\begin{figure}[!]
\centering
\includegraphics[scale=0.9]{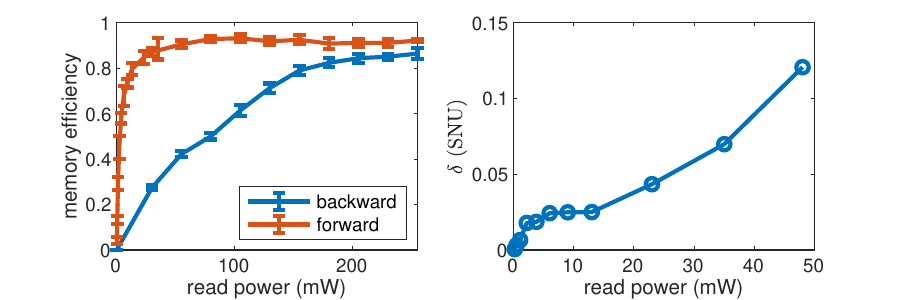}
\caption{\label{read}(a) The memory efficiencies of the forward and backward retrieval strategies vary with the power of read pulse. (b) The excess noises $\delta$ grows with the the power of read pulse.}
\end{figure}

\section{S4. Estimation of Excess Noise}
A noiseless quantum memory acts as a beam splitter, which can be described as
\begin{equation}
\hat{a}_{out} = \sqrt{\eta}\hat{a}_{in}+\sqrt{1-\eta}\hat{v}
\end{equation}
where the operator $\hat{v}$ represents the vacuum introduced by the non-unity memory efficiency. For a noisy Raman memory process, the excess noise originates from FWM amplification, which amplifies both the input state $\hat{a}_{in}$ and the introduced vacuum $\hat{v}$. To quantify the excess noise introduced in the memory process, we define the excess noise by introducing an additional thermal noise $\hat{b}_{th}$ into a noiseless beam splitter, as shown in Eq. (1) in the main text.

Using the noisy beam splitter model, we can estimate the influence of excess noise from quadrature variance data of the input and output states. Given the phase-dependent nature of the measured quadratures and the phase-independent thermal noise, we compute a phase-averaged excess noise metric:
\begin{equation}
\delta = \frac{1}{N} \sum_{i=1}^N  \frac{\langle\Delta X^2_{\text{out}}(\theta_i)\rangle - \eta \langle\Delta X^2_{\text{in}}(\theta_i)\rangle-(1-\eta)\langle\Delta X^2_v\rangle}{\langle\Delta X^2_v\rangle}.
\end{equation}
Here, $N$ is the number of phase bins over a full scan, $\theta_i$ is the phase angle of each phase bin, $\eta$ is the end-to-end memory efficiency, and $\Delta X^2_v$ is the vacuum noise. Using this method, we obtain the thermal noise contributions for the input state with different bandwidths (see Table 1 in the main text).
\end{widetext}

\bibliography{references}

\end{document}